# A Distributed Deadlock-Free Quorum-Based Algorithm for Mutual Exclusion

Mohamed NAIMI
Department of Computer Science
University of Cergy Pontoise
33, Boulevard du port
95000 Cergy-Pontoise, France

Ousmane THIARE
Department of Computer Science
UFR S.A.T
University Gaston Berger
BP. 234 Saint-Louis, Senegal

*Abstract*— **Quorum-based mutual exclusion algorithms enjoy many advantages such as low message complexity and high failure resiliency. The use of quorums is a well-known approach to achieving mutual exclusion in distributed environments. Several distributed based quorum mutual exclusion was presented. The number of messages required by these algorithms require between $3\sqrt{n}$ and $5\sqrt{n}$, where n is the size of under- lying distributed system, and the deadlock can occur between requesting processes. In this paper, we present a quorum-based distributed mutual exclusion algorithm, free deadlock. Every group is organized as a logical ring of $\sqrt{n}$ processes. A requesting process sends its request to its successor on the logical ring. When a process receives its own request after one round, it enters in the critical section. The algorithm requires $2\sqrt{n}-1$ messages.**

## I. INTRODUCTION

Distributed system is a set of processes (computers) connected by communications links. To achieve collaborative tasks by a set of processes, many distributed algorithms have been proposed. The problem of mutual exclusion is one of fundamental problem in distributed systems, which is required to, for example, update of shared object consistently. By distributed mutual exclusion, it is guaranteed that the number of processes which updates the object is at most one at any time.

In distributed systems, different processes are running on different nodes of the network and they often need to access shared data and resource, or need to execute some common events. Their uses should be consistent and so any access to them should be mutually exclusive. The portion of an event or application, where any shared components or common events are needed to be used, is the Critical Section (CS). Mutual Exclusion (ME) algorithms ensure the consistent execution of CS. As the shared memory is absent in distributed systems the solutions of the ME problem is not straight forward. Due to the enormous importance of ME and the difficulty of its solution, this is an extensive research area since last three decades. The classic algorithms for mutual exclusion that have been proposed for fixed networks can be classified in two types: centralized and distributed approaches. In the centralized solutions, a node is designated as coordinator to deliver permission to the other nodes to access their critical section while in the distributed solutions, the permission is obtained from consensus among all network nodes.

On the distributed systems, distributed mutual exclusion algorithms are mainly classified in two categories: token based [1][2][11] and permission based [3][4][5][6][9]. Permission based mutual exclusion algorithms impose that a requesting node is required to receive permissions from other nodes (a set of nodes or all other nodes). In token-based mutual exclusion algorithms, a unique token is shared among the set of nodes. The node holding the token is allowed to enter its critical section. The basic idea of token-based algorithms is simple: a node must own the unique token (sometimes cited as privilege messages) before entering the CS. So, in the best case, no communication is necessary since the token may be available locally. Otherwise, a mechanism is needed to locate the token. In [2], a spanning tree of network for locating the token is used and it shows that the average number of messages exchanged in this protocol is *O(log n)*. But token-based algorithms suffer from poor failure resiliency. In particular, if the node holding the token fails, complex token regeneration protocols must be executed.

## II. RELATED WORK

Ricart and Agrawala proposed the fair algorithm [3] that need 2(n-1) messages for a node to use the critical section. This algorithm is the first permission-based ME algorithm where a node need to collect permission from all other node for CS access. Though the algorithm is deadlock and starvation free, it is vulnerable to node and communication failure and it is expensive in communication cost too.

There is elegant class of permission-based algorithms [6] that use concept of quorums to achieve mutually exclusive access of CS. A node needs to achieve permissions from all the nodes of a quorum to access CS. Quorum based algorithms are resilient to node and communication failures and often network partitioning and usually have lower communication cost. Communication cost of these algorithms is proportional to the quorum size. Therefore these algorithms try to achieve the two goals: small quorum size with high degree of fault tolerance. Its basic idea is to collect enough permission (votes) to guarantee the mutual exclusion. The majority quorum algorithm [8] can be considered as the first algorithm of this kind, where to attain mutual exclusion, a node must obtain permission from a majority of nodes in the network. Maekawa [4], proposed an ME algorithm by imposing a logical structure on the network. In this scheme, a set of nodes is associated with each node, and this set has a nonempty intersection with all other sets corresponding to the other nodes, which guarantee the ME. The size of each of these sets is n and so the algorithm cost n order.

Garcia-Molina and Barbara [8] have properly defined the concept of quorums with the notion of coterie. A coterie is a set of sets with the property that any two members of a coterie have a nonempty intersection and the minimality property. Combining the idea of logical structures and the notion of coteries, an efficient and fault tolerant quorum generation algorithm for ME is proposed by Agrawal and Abbadi [5]. Here, the nodes form a logical binary tree which is used to generate quorums. The quorum forming in this algorithm is recursive. It can be regarded as attempting to obtain permissions from nodes along a root-to-leaf path. If the root fails, then the obtaining permissions should follow two paths: one root-to-leaf path on the left subtree and one root-to-leaf path on the right subtree. The algorithm tolerates both node failures and network partitions while in the best case incurring logarithmic costs considering the size of the network. But the cost increases with the increase of node failures.

*A. The distributed computational model*

A distributed system consists of n sites (1,2,3,...i,...,n). A distributed system is *asynchronous*, i.e., there is no common global clock. Information exchanged between processes is done by asynchronous message passing. Each communication channel is FIFO and each message sent is delivered within finite time, but there is no upper bound on message delivery time. In this section, we present the computational model for the proposed algorithm and a review of Maekawa's algorithm.

*1) Maekawa's algorithm:* In Maekawa's algorithm, a site does not request permission from all the sites, but only from a subset of sites. The sites of the system is divided into groups called quorums ($S_i$, $1 \leq i \leq n$). The quorums are constructed such as to satisfy the following conditions :

1. $\forall i \, \forall j, \, S_i \cap S_j \neq \emptyset, i \neq j, 1 \leq i,j \leq n$
2. $\forall i, node\ i \in S_i, 1 \leq i \leq n$
3. $\forall i, |S_i| = k, 1 \leq i \leq n$
4. $\forall j, node\ j\ is\ within\ k\ S_i's\ 1 \leq i,j \leq n$

Condition 1 : is a necessary condition for the $S_i$'s so that mutual exclusion requests can be resolved. Condition 2 : reduces the number of messages to be sent and received by a node. Condition 3 : means that each node needs to send and receive the same number of messages to obtain mutual exclusion (equal work).

Finally, condition 4 signifies that each node serves as an arbitrator for the same number of nodes. This ensures that each node is equally responsible for mutual exclusion (equal responsibility).

Maekawa established the following relationship between n and k defined as follows n = k(k-1)+1. Hence k can be found approximated to $\sqrt{n}$. The different types of messages used are *REQUEST, LOCKED, INQUIRY, FAILED, RELINQUISH* and *RELEASE*. Timestamps (TS) at any site *i* (where $1 \leq i \leq n$), Ts$_i$ are ordered par (H$_i$,i), containing the Lamport's logical clock [10] value H$_i$ and the site id i. Entry Section : Process i multicasts the *REQUEST* message to all the nodes in its S$_i$ including itself. The intersection nodes can send the *REQUEST* messages to any one of the districts to which they belongs. When a process j receives the *REQUEST* message, it sends *LOCKED* message to site i if it has not yet sent it to any other site from the time it received *RELEASE* message. Or else it queues the *REQUEST*.

For any node i which intends to execute its CS, the algorithm works as follows :

CS Execution : Process i executes its CS after receiving *LOCKED* message from all the nodes of its S$_i$.

Exit Section : After executing its CS, site i sends *RELEASE* message to all nodes of its S$_i$ which restores node's right to send *LOCKED* message to any other pending requests in the queue.

This basic algorithm is prone to deadlock which is handled as follows : Assume that a site j has *LOCKED* message to some site k and it later receives a *REQUEST* message from any other site i (i≠k). Then, node j sends *FAILED* to site i if TS$_k$ < TS$_i$, otherwise it sends *INQUIRY* message to site k. When such a process k receives *INQUIRY* message, it sends *RELINQUISH* message to site j if site k has received *FAILED* message from at least one site in S$_k$, and has not received new *LOCKED* message from it (after receipt of *FAILED* message).

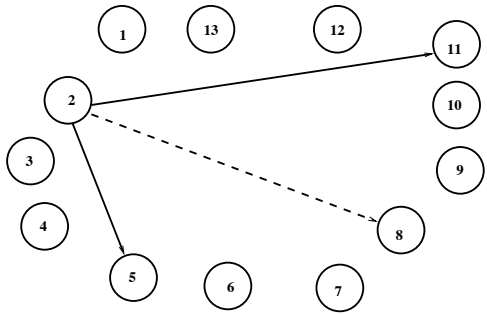

**2 requests the critical section and sends requests to processes 5, 8 and 11**

Fig. 1. Scenario 1

Example of execution: For Fig. 1, the sites are:

$S_1=\{1,2,3,4\}$
$S_2=\{2,5,8,11\}$
$S_3=\{3,5,9,13\}$
$S_4=\{4,5,10,12\}$
$S_5=\{5,1,6,7\}$
$S_6=\{6,2,9,12\}$
$S_7=\{7,3,8,12\}$
$S_8=\{8,1,9,10\}$
$S_9=\{9,4,7,11\}$
$S_{10}=\{10,2,7,13\}$
$S_{11}=\{11,3,6,10\}$
$S_{12}=\{12,1,11,13\}$
$S_{13}=\{13,4,6,8\}$

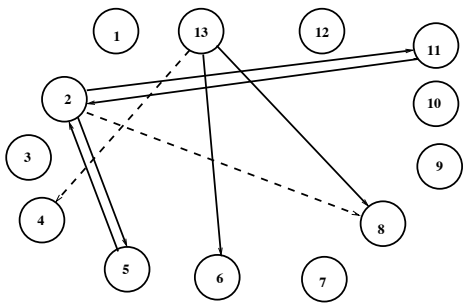

**5, 11 are locked for 2**

**13 requests the critical section and sends request to processes 4, 6 and 8.**

Fig. 2. Scenario 2

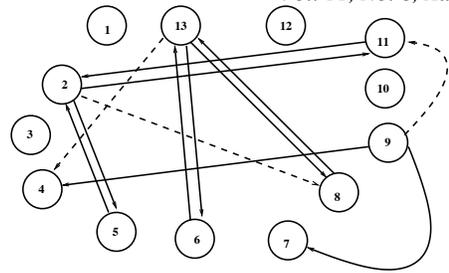

**5, 11 are locked for 2**

**6 and 8 are locked for 13**

**9 requests the critical section and sends requests to processes 4, 7 and 11**

Fig. 3. Scenario 3

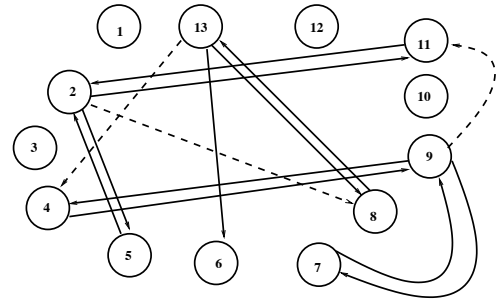

**5, 11 are locked for 2**

**6 and 8 are locked for 13**

**4 and 7 are locked for 9**

**The deadlock occurs: 2 waits 8, 9 waits 11, and 13 wait 4**

Fig. 4. Scenario 4 in presence of deadlock

### III. PRINCIPLE OF TH ALGORITHM

Each group is structured in circular ring oriented and arranged according to the identities of the process from smallest to largest.

| n=3 | n=7 | n=13 |
|---|---|---|
| $S_1=\{1,2\}$ | $S_1=\{1,3,6\}$ | $S_1=\{1,4,5,7\}$ |
| $S_2=\{2,3\}$ | $S_2=\{2,6,7\}$ | $S_2=\{2,3,7,11\}$ |
| $S_3=\{3,1\}$ | $S_3=\{3,5,7\}$ | $S_3=\{3,4,10,13\}$ |
| | $S_4=\{4,2,3\}$ | $S_4=\{4,6,11,12\}$ |
| | $S_5=\{5,1,2\}$ | $S_5=\{5,8,11,13\}$ |
| | $S_6=\{6,4,5\}$ | $S_6=\{6,7,9,13\}$ |
| | $S_7=\{7,4,1\}$ | $S_7=\{7,8,10,12\}$ |
| | | $S_8=\{8,1,3,6\}$ |
| | | $S_9=\{9,2,4,8\}$ |
| | | $S_{10}=\{10,2,5,6\}$ |
| | | $S_{11}=\{11,1,9,10\}$ |
| | | $S_{12}=\{12,3,9,5\}$ |
| | | $S_{13}=\{13,1,2,12\}$ |

We consider the groups $S_1=\{1,4,5,7\}$, $S_9=\{9,2,4,8\}$ and $S_{13}=\{13,1,2,12\}$

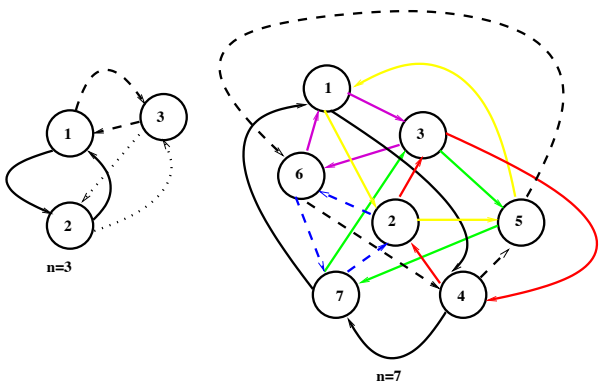

Fig. 5. Circular ordered lists

Local variable at node $P_i$ :

The variables used in the algorithm for process $P_i$ are listed below:

$Stat_i$ : indicates whether a node $P_i$ is in the **Wait**=requesting, **Ready**=in critical section or Passive=not requesting. Initially, ∀i, $Stat_i$ = **Passive** $S_i$ : set of identities of processes of $P_i$'s group.

$F_i$ : local waiting queue of nodes $P_i$. Initially $F_i=\emptyset$.

$B_i$: boolean that indicates whether a process $P_i$ is blocked or not. In the algorithm, every process uses two messages:

Req: message sent by process $P_i$ to request the critical section.
Rel: message sent by process $P_i$ to release the critical section. This message is sent to every node in $S_i$.

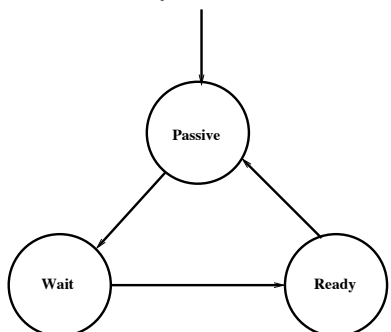

Fig. 6. States process

**Principle of the algorithm without deadlock :**

We assume that each process builds its circular list ordered $L_i$. Our algorithm do not use logical timestamps. When a node $P_i$ requests the critical section, two cases are possible: $P_i$ is placed to the waiting queue $F_i$ and there exists two cases:

**case 1**: $P_i = Min(L_i)$, then $P_i$ is placed in its local queue $F_i$, if is the head of its waiting queue, then it sends a request Req(i) to its successor in $L_i$ and waits an authorization to enter in the critical section.

**case 2**: $P_i = Max(L_i)$, then $P_i$ sends the request message Req(i) to $P_j = Min(L_i)$, and waits for authorization to enter in the critical section. When process $P_i$ release a resource, it broadcasts a message Rel(i) to all members of his group that is to say all the processes in its list $L_i$.

*A. Pseudocode of the algorithm*

**When $P_i$ requests the critical section**

$Stat_i \leftarrow$ Wait

If (($P_i = Min(S_i)$)) Then

Append($F_i$, $P_i$)

If ($P_i = Head(F_i)$) Then

Send Req($P_i$) To Succ($P_i$)

$B_i \leftarrow$ True EndIf

Else Send Req($P_i$) To Min($S_i$)

EndIf

**When $P_i$ receives Req(P)**

If ($P \not\in F_i$) Then

Append($F_i$ , P )

EndIf

If (Head($F_i$) = $P_i$) Then

State $\leftarrow$ Ready

$B_i \leftarrow$ True

Else

If (P = Head($F_i$)) Then

Send Req(P ) To Succ($P_i$)

$B_i \leftarrow$ True EndIf

EndIf

**When $P_i$ releases the critical section**

∀P⊂$S_i$ send Rel($P_i$) To P

Remove($F_i$, Head($F_i$)) $Stat_i \leftarrow$ Passive

If ($F_i \neq [\ ]$) Then

Send Req(Head($F_i$)) To Succ($F_i$)

Else

$B_i \leftarrow$ False

EndIf

**When $P_i$ receives Rel(P)**

Remove($F_i$,P)

$B_i \leftarrow$ False

If ($F_i \neq [\ ]$) Then

$B_i \leftarrow$ True

If ($P_i = Head(F_i)$)) Then

State $\leftarrow$ Ready

Else

Send Req(Head($F_i$)) To Succ(Head($F_i$))

EndIf

EndIf

*B. Example of execution*

We consider a network of 13 processes with the groups $S_1, S_2, \cdots, S_{13}$ constructed as in Section 3. We assume that processes 2, 9 and 13 request to enter the critical section. Now we illustrate the algorithm by the following scenario:

$T_1$ : Process 2 comes in its queue and waiting to become head of the queue. Once he is the head of the waiting queue, it sends a request Req(2) to his successor in his group which is process 3.

$T_2$ : Process 9 comes in its queue, it sends a request Req(9) to the smallest of its group process that is process 2.

$T_3$ : Process 13 comes in its queue, it sends a request Req(13) to the smallest of its group process that is process 1.

$T_4$ : Process 3 receives the request Req(2) and puts 2 in tail in its queue, if 2 is the head, it sends the request Req(2) to process 7, otherwise 2 remains in the queue of process 3.

$T_5$ : Process 2 receives the request Req(9) and puts 9 in its queue.

$T_6$ : Process 7 receives the request Req(2) and puts 2 in its queue and sends Req(2) to process 11.

$T_7$ : Process 1 receives the request Req(13), puts 13 in his file and becomes blocked by requesting process, process 1 forwards the Req(13) to process 2.

$T_8$ : Process 11 receives request Req(2) and puts 2 in his queue and becomes blocked for 2, it sends Req(2) to process 2.

$T_9$ : Process 2 receives the request Req(13), puts 13 in his file.

$T_{10}$ : Process 2 receives its own request Req(2) from process 11, it enters the critical section.

We have the following table :

| Process | Waiting queue | State |
|---|---|---|
| 1 | (1,3) | blocked for 13 |
| 2 | (2,9,13) | in critical section |
| 3 | (2) | blocked for 2 |
| 4 | () | blocked for 9 |
| 5 | () | blocked for 2 |
| 6 | () | free |
| 7 | (2) | free |
| 8 | () | free |
| 9 | () | requester |
| 10 | () | free |
| 11 | (2) | free |
| 12 | () | free |
| 13 | () | requester |

$T_{11}$ : Process 2 releases the critical section, and broadcasts a message Rel(2) to all members in its group $S_2$ 3,7,11. The process 2 sends the blocked request of process 9 to process 4.

$T_{12}$ : Process 4 receives Req(9) from process 2, it puts it in its file, and forwards it to process 8.

$T_{13}$ : Process 8 receives Req(9) from process 4, it puts it in its file and forwards it to process 9.

$T_{14}$ : Process 9 receives its own request Req(9), enters its own queue. Process 9 is at the head of its file, it becomes blocked and enters in its critical section.

$T_{15}$ : Process 9 releases the critical section and broadcasts the message Rel(9) to all members of his group, i.e the processes 2,4,8.

$T_{16}$ : Process 2 receives the message Rel(9) from process 9, it removes the process 9 from its file, and sends the request of process 13 to process 12.

$T_{17}$ : Process 12 receives the message Req(13) from 2, it puts the process 13 in its queue and sends Req(13) to 13.

$T_{18}$ : Process 13 receives its own request Req(13), enters its own queue. Process 13 is at the head of its file, it becomes blocked and enters in its critical section.

We have the following table :

| Process | Waiting queue | State |
|---|---|---|
| 1 | (13) | blocked for 13 |
| 2 | (13) | blocked section |
| 3 | () | free |
| 4 | () | free |
| 5 | () | free |
| 6 | () | free |
| 7 | () | free |
| 8 | () | free |
| 9 | () | free |
| 10 | () | free |
| 11 | () | free |
| 12 | (13) | blocked for 13 |
| 13 | (13) | in critical section |

IV. PROOF AND CORRECTNESS

*A. Mutual exclusion*

Mutual exclusion is achieved when no pair of processes is ever simultaneously in its critical section. For any pair of processes, one must leave its critical section before the other may enter.

*Theorem 4.1:* The proposed algorithm ensures the mutual exclusion property.

*Proof:* Assume the contrary, that more than one node are simultaneously in the critical section. Suppose that two application processes $P_i$ and $P_j$ ($i \neq j$) in different groups are in the critical section simultaneously. Let $S_i$ and $S_j$ be groups that $P_i$ and $P_j$ belong respectively. Because any two groups have non-empty intersection, we have $S_i \cap S_j \neq \emptyset$ and let $P_k$ be a process in the intersection. Since $P_k$ never grants permission for more that one group at a time, $P_i$ and $P_j$ cannot be granted by $P_k$ simultaneously. This is a contradiction.

### B. Deadlock and starvation freedom

*1) Deadlock freedom:* Maekawa's algorithm can deadlock because a process is exclusively locked by other processes and requests are not prioritized by their timestamps.

*Proof:* Deadlock handling in [4] requires three types of messages: failed, inquire and yield.

Deadlock could occur for a set of processes if they were each involved in a circular wait. A circular wait could occur if each of the processes $P_i$ in the cycle is blocked at the waiting queue located at process $P_j$, and is yet to receive a request message from the successor process in the cycle and no there are no request in transit which are destined for any of these processes. Assume, by way of contradiction, that this is the case. Then each process in the circular wait has delayed sending a request message to its predecessor process in the cycle. A processes $P_i$ will only defer sending a request to a process $P_j$. Thus, to achieve a deadlock, each process in the circular wait must be blocked by its predecessor process in its group, which is impossible. Therefore, the algorithm is deadlock-free.

*2) Starvation freedom:* Starvation occurs when a few processes repeatedly execute their critical section while other processes wait indefinitely. Assume, by way of contradiction, that process $P_j$ has been repeatedly executing its critical section while process $P_i$ has been waiting to enter in its critical section.

The groups of processes are organized as a logical ring of processes, and every process knows its successor on the ring. Every process uses a local waiting queue to store the pending requests.

*Theorem 4.2:* Every request process enter in the critical section during a bounded delay.

*Proof:* Every process receives, at most one, request from every process in its group. Every request is stored in its waiting queue for a bounded delay.

By examining the algorithm, when process releases its criticalsection, it sends a release message to all processes in its group.

when a process receives a release message, it removes the request placed at the head of its waiting queue. At most $\sqrt{n}$ request are placed in a waiting queu before any request. A request transits by $\sqrt{n}$ processes of its group.

### C. Message complexity

The message complexity of a distributed mutual exclusion algorithm is the number of messages exchanged by a process per critical section.

*Theorem 4.3:* Message complexity of the proposed algorithm is $2\sqrt{n}$ in the best case and $O(3|S|)$ in the worst case, where $|S|$ is a quorum size that the algorithm adopts.

*Proof:* In the best case, two types of messages (Req, Rel) are exchanged between application process and each management process in a quorum. Thus, message complexity is $2|S|$ in the best case, where $|S|$ is a quorum size that the algorithm adopts. Outline of the scenario of the worst case is as follows. A process $P_i$ send a request message Req to $P_j$ in the group $S_i$, but $P_i \neq \min(S_i)$ and $P_i \neq \max(S_i)$. In addition to the best case, additionally one (1) message is exchanged, we have the bound $|S| + 2|S| = O(3|S|)$.

## V. CONCLUSION

Quorum-based mutual exclusion is an attractive approach for providing mutual exclusion in distributed systems due to its low message complexity and high resiliency. After the first quorum-based algorithm [4] was proposed by Maekawa more than a decade ago, many algorithms [3][4][5][6][9] have been proposed to construct different quorums to reduce the message complexity or increase the resiliency to site and communication failures. Some researchers also propose schemes for constructing delay-optimal quorums to reduce the average message delay. However, all these quorum-based algorithms depend on Maekawa's algorithm to ensure mutual exclusion and they all have high synchronization delay (2T).

We have presented a very simple free deadlock distributed mutual exclusion algorithm based on quorum principle. Every group is structured to ordering circular list, and every process is am smallest or the biggest of his group. The request message sends by a requesting process, visits all processes according to the order of its list. Every critical section execution, requires at least $2\sqrt{n}$ messages where n is the number of processes in the network.

### REFERENCES

[1] S. Banerjee, and P. Chrysanthis, "A New Token Passing Distributed Mutual Exclusion Algorithm," Proceedings of the 16[th] ICDCS, pp. 717-724, 1996.

[2] M. Naimi, and M. Trehel, "How to detect a failure and regenerate the Token in the Log(n) distributed algorithm for mutual exclusion," LNCS 312, Amsterdam, 1987.

[3] G. Ricart, and A. K. Agrawala, "An Optimal Algorithm for Mutual Exclusion in Computer Networks," Communications of the ACM, Vol. 24, No. 1, pp. 9-17, 1981.

[4] M. Maekawa, "A $\sqrt{n}$ Algorithm for Mutual Exclusion in Decentralized Systems," ACM Trans. Computer Systems, vol. 3, No. 2, pp. 145-159, 1985.

[5] D. Agrawal, and A. El Abbadi, "An Efficient and Fault-Tolerant Solution for Distributed Mutual exclusion," ACM Trans. On Computer systems, Vol. 9, No. 1, pp. 1-20, 1991.

[6] C. Saxena, J. Rai, "A survey of permission-based distributed mutual exclusion algorithm," Elsevier Science Publisher B. V., Vol. 25, No. 2, pp. 159-181, 2003.

[7] R. H. Thomas, "A majority consensus approach to concurrency control," ACM Trans. On Database System, Vol. 4, No. 2, pp. 180-209, 1979.

[8] H. Garcia-Molina, and D. Barbara, "How to assign votes in a distributed system," Journal of the ACM, Vol. 32, No. 4, pp. 841-860, 1985.

[9] L. Lamport, "Time, clocks, and the ordering of events in a distributed system", Communications of the ACM, Vol. 21, No. 7, pp. 558-565, 1978.

[10] R. Atreya, and N. Mittal, "A quorum-based group mutual exclusion algorithm for a distributed system with dynamic group set", In IEEE Trans. On Parallel and Distributed Systems, Vol. 18, No. 10, 2007.

[11] I. Suzuki, and T. Kasami, "A distributed mutual exclusion algorithm," ACM Trans. On Computer Systems, Vol. 3, No. 4, pp. 344-349, 1985.

AUTHORS PROFILE

**Mohamed Naimi.** Received a PhD in computer science (Distributed systems) from the university of Franche-Comté Besancon, France. He is a Full Professor in the University of Cergy-Pontoise. He has been author and co-author of published papers in several journals and recognized international conferences and symposiums.

**Ousmane Thiare.** Received a PhD in computer science (Distributed systems) at 2007 from the university of Cergy Pontoise, France. He is an Associate Professor in Gaston Berger University of Saint-Louis Senegal. He has been author and co-author of published papers in several journals and recognized international conferences and symposiums.